\documentclass[twocolumn,nopacs,preprintnumbers,amsmath,amssymb]{revtex4}
\usepackage{graphicx}
\begin{document}

\title{Effect of magnetic field and chemical potential   on the RKKY interaction in the $\alpha$-${\cal T}_3$ lattice  }

\author{ Oleksiy Roslyak$^{1}$,  Godfrey Gumbs$^{2}$, Antonios Balassis$^{1}$, Heba Elsayed$^{1}$  }

\affiliation{$^1$ Department of Physics \& Engineering Physics, Fordham University,  441 East Fordham Road,  Bronx, NY  10458 USA}
\affiliation{$^{2}$Department of Physics and Astronomy, Hunter College of the
City University of New York, 695 Park Avenue, New York, NY 10065, USA}


\begin{abstract}

The interaction energy for the indirect-exchange or Ruderman-Kittel-Kasuva-Yosida (RKKY) interaction between magnetic spins localized on lattice sites of the  $\alpha$-${\cal T}_3$ model is calculated using linear response theory. In this model, the  $\texttt{AB}$-honeycomb lattice structure is supplemented with $\texttt{C}$ atoms at the centers of the hexagonal lattice. This introduces a parameter $\alpha$ for the ratio of the hopping integral from hub-to-rim and that around the rim of the hexagonal lattice. A valley and $\alpha$-dependent  retarded Green’s function matrix   is used to form the susceptibility. Analytic and numerical results are obtained for undoped $\alpha$-${\cal T}_3$, when the chemical potential is finite and also  in the presence of an applied magnetic field. We demonstrate the anisotropy of these results when the magnetic impurities are placed on the $\texttt{A,B}$ and  $\texttt{C}$ sublattice sites.  Additionally, comparison of the behavior of the susceptibility of $\alpha$-${\cal T}_3$ with graphene shows that there is a phase transition at $\alpha=0$.

\end{abstract}
\maketitle

\section{Introduction}\label{sec1}

\begin{figure}[h!]
\centering
\includegraphics[width=0.40\textwidth]{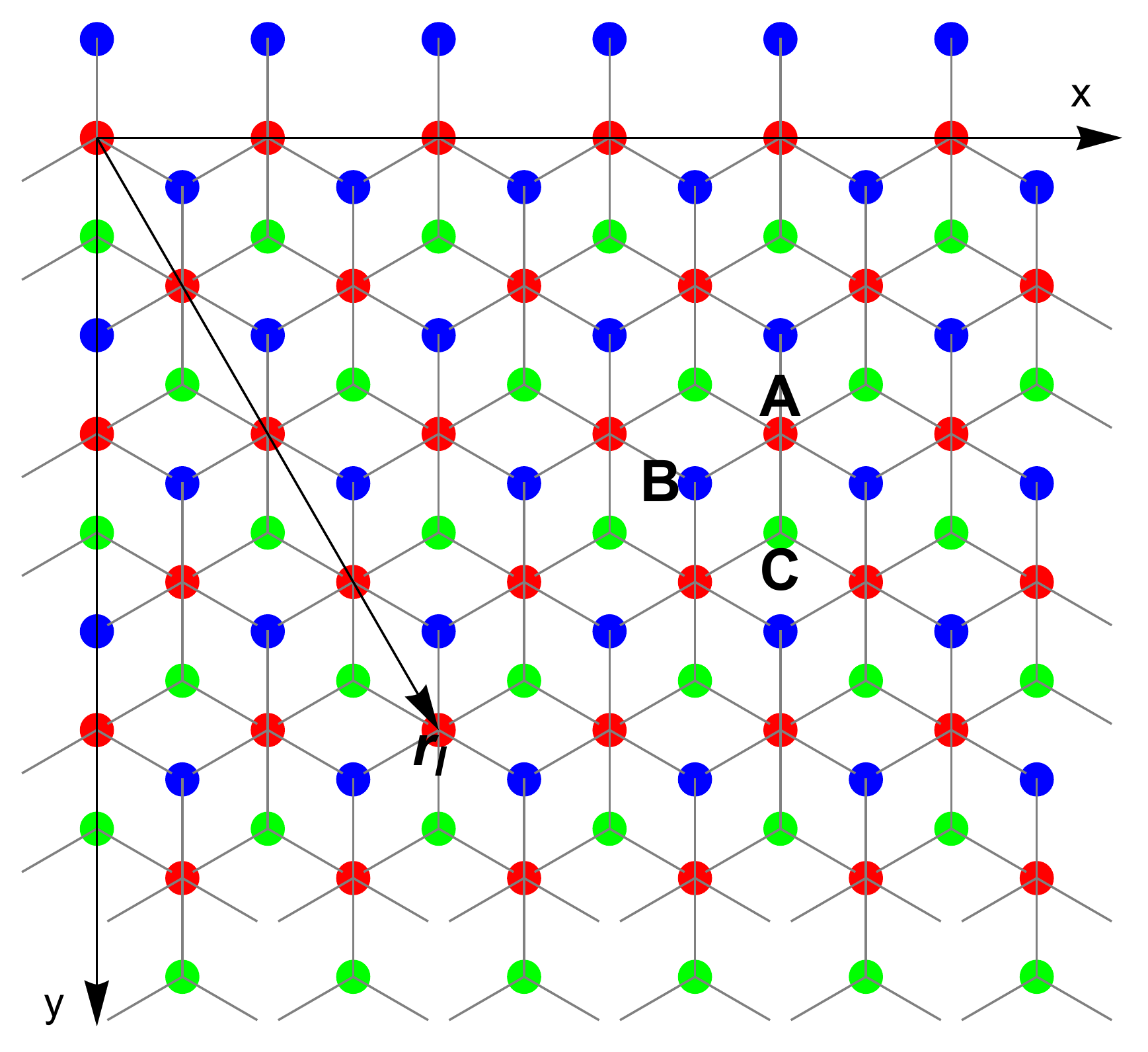}
\caption{(Color online)\ Lattice sites of the $\alpha-\mathcal{T}_3$ model.  The ``rim" atoms are labeled $\texttt{A}$ and $\texttt{B}$ whereas $\texttt{C}$ is a ``hub" atom.}
\label{fig:1}
\end{figure}

An effective single-particle model Hamiltonian representing an electronic crystal has been recently constructed  to represent  the low-lying Bloch band of the $\alpha$-${\cal T}_3$ lattice.  (For a review of artificial flat band systems, see Ref.\  \cite{Review}.) The electronic properties of this material have come under growing scrutiny  for a number of important reasons which are  fundamental and technological   \cite{f1,f2,f3,f4,f5,BNRef5,BNRef6,BNRef6A,BNRef6B,f6,f7,f8,f9,RKKY,BNRef8,BNRef9,BNRef10,t1,t2,t3,Dey}.    The potential tunability of these materials ranging from their optical and transport properties to their response to a uniform magnetic field and varying chemical potential presents researchers with the opportunity to investigate new materials.  Regarding their fabrication, it was suggested in \cite{f1} that an $\alpha$-${\cal T}_3$ lattice may be constructed  with the use of cold fermionic atoms  confined to an optical lattice with the help of three pairs of laser beams for the optical dice ($\alpha=1$) lattice \cite{21}. Jo, et al. \cite{BNRef6A}  successfully fabricated a two-dimensional kagome lattice consisting of ultracold atoms by superimposing a triangular optical lattice on another one commensurate with it, and  generated by light at specified wavelengths.   The $\alpha$-${\cal T}_3$ and kagome lattices are related in that they both have flat bands as well as Dirac cones at low energies.  In modeling this structure, an $\texttt{AB}$-honeycomb lattice like that in graphene is combined with $\texttt{C}$ atoms at the centers of the hexagonal lattice as depicted in Fig.\ \ref{fig:1}.  Consequently, a parameter $\alpha$ is introduced to represent the ratio of the hopping integral between the hub and the rim ($\alpha t$) to that around the rim ($t$) of the hexagonal lattice. When one of the three pairs of laser beams is dephased,  it is proposed in \cite{21} that this could allow the possible variation  of the hopping parameter over the range $0<\alpha <1$.

\medskip
\par

Interestingly, it would be informative to explore how the optical and transport properties of $\alpha$-${\cal T}_3$ systems are affected by defects.  These include substituting impurities or guest atoms in a hexagonal lattice with fermionic host atoms.  In  this way, one could effectively manipulate the fundamental properties which are inherent to the $\alpha$-${\cal T}_3$ system.  The guest atoms could be added to their hosts by chemical vapor deposition (CVD) or discharge experiments. With doping, the $\texttt{A}$ and $\texttt{B}$ sublattices are no longer equivalent since the $\pi$ bonding on these lattices may be seriously distorted and this causes  significant modification of the physical properties, including the energy band structure with a deviation from the original Dirac cone and flat band. However, at low doping  ($< 1.5\%$), the low-energy  portion of the band structure is only slightly affected.  But, we emphasize that the doping configuration and concentration in general create unusual band structures with feature-rich and unique properties.  

\medskip
\par

Oriekhov  and   Gusynin \cite{RKKY} took the first step of investigating  the role played by the sea of background $\alpha$-${\cal T}_3$-fermions on the indirect exchange interaction between a pair of spins localized on lattice sites.  Local moments like these may occur near extended defects. The doping giving rise to the presence of these spins was assumed to have such a low concentration that  the energy dispersion is unaltered, as well as there is no change to the zero band gap.  Specifically, these authors \cite{RKKY} were interested in this effect of doping and temperature on the Ruderman-Kittel-Kasuva-Yosida (RKKY) or indirect-exchange coupling as it was discussed for different types of two-dimensional (2D) materials by others \cite{O1,O2,O3,Nanotube,WANG}  between spins via the host conduction electrons of free standing monolayer  graphene, \cite{1,20,O4,O5,O6,O7,O8,O9,O10,O11,Fertig1} and biased single-layer silicene \cite{RKKYsilicene}.  In this paper, we continue the investigation in \cite{RKKY} by calculating the effect of a uniform magnetic field and variable chemical potential on the RKKY interaction of $\alpha$-${\cal T}_3$.  It is worthwhile getting a better understanding of the behavior of this topic since one could exploit the RKKY interaction to determine spin ordering as excitations near the Fermi level are in part governed by the indirect exchange interaction between local magnetic moments \cite{GG1,Roldan,GG2}.

\medskip
\par

The outline of the rest of this paper is as follows.  In Sec.~\ref{sec2}, we present the low-energy $\alpha$-${\cal T}_3$ model Hamiltonian and derive the lattice Green's functions for the intrinsic case without magnetic field.  Section \ref{sec3} is devoted to a calculation of the indirect exchange coupling between a pair of  impurities in the absence and presence of doping and uniform magnetic field. We shall represent RKKY interaction energy as a Hadamard product of three matrices: valley matrix, $\alpha-$matrix and distance matrix.
Our numerical results for the $\alpha$-dependent exchange interaction are given in Sec.\ \ref{sec4}.  We demonstrate that the spin susceptibility for the $\alpha$-${\cal T}_3$ model is different in nature from that for graphene thereby signaling a magnetic phase transition at $\alpha=0$.  We analyze the behavior of the spin susceptibility at low magnetic field and when the doping is high in Sec.\ \ref{sec4}. We conclude with a summary in Sec.\ \ref{sec5}.

\medskip
\par

\section{ The $\alpha$-${\cal T}_3$ model Hamiltonian and lattice Green's functions} 
\label{sec2}


The goal of this section is to introduce the lattice specific Green's functions which are essential for calculating RKKY interactions.
Throughout the paper we use the conventions: bold capitalized letters stand for $3 \times 3$ matrices (or $3\times 1$ vectors); tilded quantities are dimensionless.
Absent the magnetic field, the energy spectrum can be derived from the  low-energy Hamiltonian  at the $K$ and $K^\prime$ points:

\begin{equation}
\mathbf{H} =\left(
\begin{array}{ccc}
  0   &  f_{\lambda,\mathbf{k}} \cos \phi  & 0 \\
  f^\ast_{\lambda,\mathbf{k}}  \cos \phi   & 0   &  f_{\lambda,\mathbf{k}} \sin \phi  \\
 0 & f^\ast_{\lambda,\mathbf{k}} \sin \phi   & 0    \\
\end{array}
\right) \ ,
\label{eqn:1}
\end{equation}  
where  $f_{\lambda,\mathbf{k}}=\lambda\epsilon_k \  e^{-i\lambda\theta_k}$ 
with $\epsilon_k=\hbar v_Fk$; $\lambda =\pm 1$ stands for the valley index at the $K$ and $K^\prime$ points located at $\left(\lambda \frac{4\pi}{3\sqrt{3}a},0\right)$ with $a$ being conventional graphene carbon-carbon distance and $v_F$ stands for the Fermi velocity. The angle between $\mathbf{k}$ and the $x-$axis is given by $\theta_k$ yielding $k_x/|{\bf k}|=\cos\theta_k$, $k_y/|{\bf k}|=\sin\theta_k$.
The rows and columns of the Hamiltonian are labeled by the ($\texttt{A,B,C}$) lattice indices indicated in Fig.\ref{fig:1}.
\par
The diagonalization of the Hamiltonian \eqref{eqn:1} is readily carried out analytically to obtain the eigenenergies $\epsilon_{sk} = s \epsilon_{k}$ and eigenstates $\mathbf{\Psi}_{s,\lambda,\mathbf{k}} \left({\mathbf{r}}\right) = \mathbf{\Psi}_{s,\lambda,\mathbf{k}} \frac{e^{i{\bf k}\cdot {\bf r}}}{\sqrt{\mathcal{A}}}$. Here we have introduced a band index $s$ so that $s=0$ stands for the flat band, $s=\pm1$ indicates conduction/valence band respectively. The area of the hexagonal unit cell with side $a$ is denoted by ${\cal A}=\frac{3\sqrt{3}a^2}{2}$. 
For the flat band the normalized eigenstates are:
\begin{equation}
\mathbf{\Psi}_{\lambda,0,\mathbf{k}} = \left(
\begin{array}{ccc}
  e^{ -i \lambda \theta_k} \sin\phi \\
0\\
- e^{ i  \lambda\theta_k} \cos\phi 
\end{array} \right).
\label{eqn:3}
\end{equation}
The wave functions for the conduction/valence are
\begin{equation}
\mathbf{\Psi}_{\lambda,\pm 1,\mathbf{k}}=\frac{1}{\sqrt{2}}\left(
\begin{array}{ccc}
\lambda  e^{- i  \lambda \theta_k}      \cos\phi \\
\pm 1\\
\lambda e^{i\lambda \theta_k}      \sin\phi 
\end{array} \right).
\label{eqn:4}
\end{equation}
The eigenvalues for the unmodulated lattice  are the same near the $K$  and $K^\prime$ points but the valence and conduction bands near these points differ.

\par 
We define the Green's functions as the elements of an inverse matrix involving the energy difference with the Hamiltonian \eqref{eqn:1} as:
\begin{gather}
\mathbf{G}({\bf k},E;\lambda;\phi)=\left[{(E+i0^+) {\bf I}-\mathbf{H}}\right]^{-1}=\\
\notag=
\left(
\begin{array}{ccc}
 G_{\texttt{AA} } &  G_{\texttt{AB} } & G_{\texttt{AC} } \\
 G_{\texttt{AB} }^\ast  & G_{\texttt{BB} } &  G_{\texttt{BC} } \\
 G_{\texttt{AC} }^\ast &G_{\texttt{BC} }^\ast  & G_{\texttt{CC} } \\ 
\end{array}
\right) \ .
\end{gather}
Here $\mathbf{I}$ is the unit matrix and the  replacement $E\to E+i0^+$ guaranties retarded nature of the Green's functions.
The direct diagonalization of the Green's tensor yields:
\begin{gather}
\label{eqn:GG1}
\mathbf{G}({\bf k},E;\lambda;\phi)=\\
\notag
\notag
= \frac{1}{||\mathbf G||}
\left({
\begin{array}{ccc}
E^2-\epsilon_k^2\sin^2\phi & E f_{\lambda,\mathbf{k}}  \cos \phi & f^2_{\lambda,\mathbf{k}}\sin(2\phi)/2\\
{} & E^2 & E  f_{\lambda,\mathbf{k}}   \sin\phi \\
{} & {} & E^2-\epsilon_k^2\cos^2\phi\\
\end{array}
}\right) \ ,
\end{gather}
with the determinant given by $||\mathbf{G}({\bf k},E)||= E\left(E^2 -\epsilon_k^2  \right)$.
Alternative derivation of Eq.\eqref{eqn:GG1} based on the eigenfunction decomposition is given in Appendix \ref{AP:1}.
\par
Clearly, the Green's function matrix is Hermitian and we observe that $G_{\texttt{BB}}({\bf k},E;\lambda;\phi)=G_{\texttt{AA}}({\bf k},E;\lambda;\phi=0)$ is the only element of the Green's function matrix which does not depend on $\phi$. Consequently, this would lead to the RKKY interaction  between spins on the $\texttt{B}$ site to be unaffected when $\phi$ is varied.

Now, defining the Fourier transform of the {\em total\/} Green's function at the two valleys, upon shifting to the Dirac points with ${\bf k}\to {\bf k}+\lambda {\bf K}$, we obtain the components in the real space:

\begin{gather}
\label{defn1}
G_{\mu\nu}({\bf r}_{ll' },E;\phi)=\\
\notag=\frac{{\cal A}}{(2\pi)^2}\sum_{\lambda=\pm 1} \int_{B.Z.} d^2{\bf k}\ G_{\mu\nu}({\bf k},E;\lambda;\phi)\texttt{e}^{i({\bf k}+\lambda{\bf K})\cdot{\bf r}_{ll'}},
\end{gather}
where the integration over the wave vector ${\bf k}$ is carried out over the Brillouin zone (B.Z.) and we have used $\mathbf{r}_{ll'} = \mathbf{r}_{l}-\mathbf{r}_{l'}$. After some straightforward algebra (see Appendix \ref{AP:2}) we obtain the Green's function tensor as a Hadamard product 
\begin{equation}
\label{EQn:GreensFunction}
    \mathbf{G}({\bf r}_{ll'},E;\phi)=  \frac{{\cal A}}{\pi a^2 E} \mathbf{V}^{1/2} \circ \mathbf{\Phi}^{1/2} \circ \mathbf{R}^{1/2},
\end{equation}

where 
the valley matrix is given by

\begin{gather*}
\mathbf{V}^{1/2}\left({\mathbf{r}_{ll'}}\right)=\\
\notag
\left({
   \begin{matrix}
   \cos{\left({\mathbf{K}\cdot \mathbf{r}_{ll'}}\right)} && \sin{\left({\mathbf{K}\cdot \mathbf{r}_{ll'}-\alpha_{ll'}}\right)} && \cos{\left({\mathbf{K}\cdot \mathbf{r}_{ll'}-2\alpha_{ll'}}\right)}\\
   {} && \cos{\left({\mathbf{K}\cdot \mathbf{r}_{ll'}}\right)} && \sin{\left({\mathbf{K}\cdot \mathbf{r}_{ll'}-\alpha_{ll'}}\right)}\\
   {} &&  {} &&  \cos{\left({\mathbf{K}\cdot \mathbf{r}_{ll'}}\right)}
   \end{matrix}
}\right),
\end{gather*}
the $\alpha$ (or equivalently $\phi$) dependent matrix is 
\begin{gather*}
    \mathbf{\Phi}^{1/2}\left({\phi}\right) =
\left({
\begin{matrix}
\cos^2\phi &&  \cos \phi && \sin\left({2\phi}\right)\\
{} && 1 && \sin^2\phi\\
{} && {} && \sin^2\phi
\end{matrix}
}\right),
\end{gather*}
and the position  and energy dependent distance matrix is 
\begin{gather*}
    \mathbf{R}^{1/2}\left(E,{r_{ll'}}\right) =\\
    \notag
    =
\omega^2 \left({
\begin{matrix}
-K_0\left({-i\omega r}\right) &&  -i K_0\left({-i\omega r}\right) && \frac{1}{2}K_2\left({-i\omega r}\right)\\
{} && -K_0\left({-i\omega r}\right) && -i K_0\left({-i\omega r}\right)\\
{} && {} && -K_0\left({-i\omega r}\right)
\end{matrix}
}\right).
\end{gather*}

For convenience, we have introduced the following notation 
normalized energy $\omega=E/E_0$ with $E_0=\hbar v_F a^{-1}$, dimensionless length $r=r_{ll'} a^{-1}$ with $a$ denoting the $ \texttt{AB}$ separation on the lattice depicted in Fig.~\ref{fig:1}.

\section{Indirect Exchange Interaction between two magnetic impurities }
\label{sec3}

We now consider two magnetic impurities having spins ${\bf S}_1$ and ${\bf S}_2$ occupying the lattice sites ${\bf r}_{l}$ and ${\bf r}_{l^\prime},$, respectively. The effective RKKY exchange interaction energy for this pair of spins in the sea of Dirac electrons is within linear response theory given in the Heisenberg form as \cite{21,1,20}

\begin{equation} 
E_{\mu\nu}(r_{ll'};\phi)=\frac{\lambda_0^2\hbar^2}{4}\chi_{\mu\nu}\left(r_{ll'};\phi  \right)\  {\bf S}_1 \cdot {\bf S}_2\ ,
\end{equation}
where $\lambda_0$ is the short-range exchange interaction between the impurity spins and the $\alpha$-${\cal T}_3$  electrons, and $\chi_{\mu\nu}\left(r_{ll'};\phi  \right)$ is the free-particle  charge density sublattice susceptibility which depends on which lattice site $\mu,\nu = \texttt{A},\texttt{B},\texttt{C}$ the impurity spins are positioned at.

\subsection{Zero Fermi energy and magnetic field}

\begin{figure}[h!]
\centering
\includegraphics[width=0.45\textwidth]{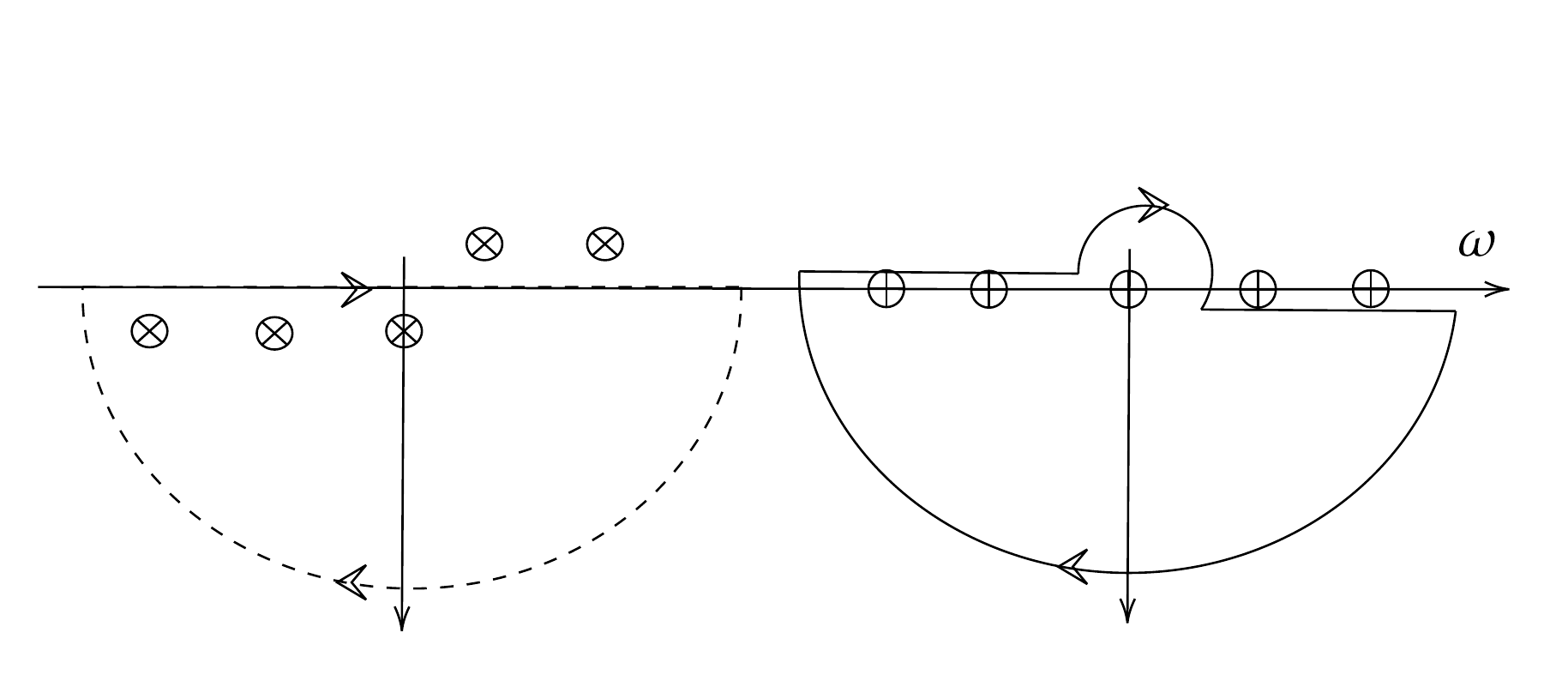}
\caption{Left panel: conventional contour spanning lower half-plane assuring the retarded form of the Green's functions. The poles shifted by $\pm i 0^{+}$ are indicated by $\otimes$. The arrows indicate the direction of the contour integral. Right panel: the integration contour $\mathcal{C}$ used in Eq.  \eqref{EQn:1} and in the calculation of the poles contribution.  The symbols $\oplus$ indicate the location of the poles of the Green's functions in  Eq.\eqref{EQn:GreensFunction}, i.e.,  $\omega\left({\omega^2-q^2}\right)=0$. The central semicircle radius is taken to be small.}
\label{fig:2}
\end{figure}

For undoped $\alpha$-${\cal T}_3$ ,  the Fermi energy is at $E_F=0$  so that we obtain the matrix of spin-dependent sublattice susceptibility \cite{1,20}:

\begin{gather}
	\label{EQn:1}
    \chi_{\mu \nu}\left({r_{ll'}; \phi}\right) = -\frac{2}{\pi}\int \limits_{-\infty}^{0} dE\
				\texttt{Im}\left[{G^2_{\mu \nu}\left({E}\right)}\right]=
   \\
   \notag
     -\frac{1}{\pi}\int \limits_{-\infty}^{0} dE\
    \texttt{Im}\left[{G^2_{\mu \nu}\left({E+i0^{-}}\right)}\right]-\frac{1}{\pi}
    \int \limits_{0}^{\infty} dE\
    \texttt{Im}\left[{G^2_{\mu \nu}\left({E+i0^{+}}\right)}\right]
    \\
    \notag
    =  -\frac{E_0}{\pi} \oint \limits_{C} d \omega\   \texttt{Im}\left[{G^2_{\mu\nu}\left({\omega}\right)}\right].
\end{gather}
The Green's functions $G_{\mu\nu}$, are given in the  preceding section. The integration contour $C$ is shown in Fig.~\ref{fig:2}.
Its path assures the retarded form of the Green's function and captures the right poles without shifting them on the imaginary axis by $\pm i 0^{+}$.

\medskip
\par
 
It is important to emphasize the reason why we have chosen to execute the integration in the extended complex plane instead of  the direct approach described in Ref.\ \cite{RKKY}. As it was indicated by the authors, obtaining an analytic expression is nontrivial since some integrals diverge at the upper limit. Their proposed regularization scheme leads to a set of inverse Mellin transforms (MT). Although the MTs are suitable for studying physically relevant asymptotic, such as the system's long range interactions, it leaves out some important questions regarding the zero band contribution.  Those authors claim that $\chi_{\texttt{A} \texttt{C}}\sim -\delta \left({E_F}\right)$ divergence due to this band contribution.  In the following, we demonstrate that such divergence depends on the order the limits are taken.  
Our contour regularization
corresponds, in fact, to the order of limits first $T \rightarrow 0$ and then $E_F \rightarrow 0$. We do not consider the other order of limits recovering $1/T$ behavior as in Ref.~\cite{RKKY}.
We obtain exact analytic expressions for all interaction ranges.

\medskip
\par
Our calculations show that the  susceptibility can be expressed in the following closed form analytic  expression
\begin{equation}
\chi_{\mu \nu} =
\left({\frac{3 \sqrt{3}}{2 \pi E_0 }}\right)^2 E_0 V_{\mu \nu}
\left({\mathbf{r}_{ll'}}\right)
\tilde{\chi}_{\mu \nu} \left({r_{ll'}; \phi}\right)   \ ,
\end{equation}
where a new valley matrix is given by $ \mathbf{V} = \mathbf{V}^{1/2} \circ \mathbf{V}^{1/2}$.
The rest of the section is focused on the dimensionless matrix elements $\tilde{\chi}_{\mu \nu}$. 
\par
We start with the RKKY interaction between the $\texttt{A}$, and $\texttt{A}$ sites of the lattice. Setting the Green's function \eqref{EQn:GreensFunction} into the general susceptibility expression \eqref{EQn:1} yields

\begin{gather*}
\label{EQ:chiAA}
  \tilde{\chi}_{\texttt{A}\texttt{A}}  =\\
  \notag
  =
  \mathcal{H}_0
  \left({-\frac{1}{\pi} \oint \limits_{C} d \omega \  \texttt{Im}\left[{
  \frac{\omega^2-q_1^2 \sin^2\phi}{\omega \left({\omega^2 -q^2_1}\right)}
  \frac{\omega^2-q_2^2 \sin^2\phi}{\omega \left({\omega^2 -q^2_2}\right)}
  }\right]
  }\right)
\\
\notag
=\mathcal{H}_0 \left(      {
\frac{1}{q_1+q_2}+
\frac{q_1 \sin^2\phi}{q_2 \left({q_1+q_2}\right)}+
\frac{q_2 \sin^2\phi}{q_1 \left({q_1+q_2}\right)}+
\frac{\sin^4\phi}{q_1+q_2}   }\right.
\\
\notag
- \left.{\frac{q_1 \sin^4\phi}{q_2 \left({q_1+q_2}\right)}
-\frac{q_2 \sin^4\phi}{q_1 \left({q_1+q_2}\right)}}\right) \  .
\end{gather*}
Here,  $q = k a$ is the normalized wave vector. The contour integral in the above equation is calculated using residues at $\omega = -q_1, \; -q_2$. Notice that the residue at $\omega = 0$ is equal to to zero. Therefore, the exact shape of the contour around this pole is irrelevant. The double Hankel transform operator is defined as
\begin{gather*}
    \mathcal{H}_p
    =
    \int \limits_{0}^{\infty} dq_1 \  q_1 J_p \left({q_1 r_{ll'}}\right) 
    \int \limits_{0}^{\infty} dq_2 \  q_2 J_p \left({q_2 r_{ll'}}\right).
     \end{gather*}
Its action yielded a rather simple expression given by

\begin{equation}
    \tilde{\chi}_{\texttt{A}\texttt{A}}  =
    \frac{1}{2 \pi r^3_{ll'}}\frac{\pi^2}{8}\left({1-\sin^2\phi}\right)^2 \ .
\end{equation}
In similar fashion, we have calculated the susceptibility between the $\texttt{A}$ and $\texttt{B}$ sites of the lattice as

\begin{eqnarray}
\label{EQ:chiAB}
  \tilde{\chi}_{\texttt{A}\texttt{B}}   &=&
  \mathcal{H}_1
  \left({-\frac{1}{\pi} \oint \limits_{C} d \omega\texttt{Im}\left[{
  \frac{q_1 \cos\phi }{\left({\omega^2 -q^2_1}\right)}
  \frac{q_2 \cos \phi}{\left({\omega^2 -q^2_2}\right)}
  }\right]
  }\right)
  \nonumber\\
& = & \mathcal{H}_1 \left(     {
-\frac{1}{q_1+ q_2}
}\right)=
-\frac{3 \pi^2/8}{2 \pi r_{ll'}^3}\cos^2\phi \  .
\end{eqnarray}
Also, for that between the $\texttt{A}$ and the $\texttt{C}$ sites of the lattice, we obtain

\begin{eqnarray}
  \tilde{\chi}_{\texttt{A}\texttt{C}}  &=&
  \mathcal{H}_2
  \left({-\frac{1}{\pi} \oint \limits_{C} d \omega\  \texttt{Im}\left[{
  \frac{q^2_1 \sin{\left({2 \phi}\right)}}{2 \omega \left({\omega^2 -q^2_1}\right)}
  \frac{q^2_2 \sin{\left({2 \phi}\right)}}{2 \omega \left({\omega^2 -q^2_2}\right)}  }\right]
  }\right)  
\nonumber\\
&=&   \mathcal{H}_2 \left({
-\frac{\sin^2{\left({2 \phi}\right)}}{4 \left({q_1+q_2}\right)}
-\frac{q_1\sin^2{\left({2 \phi}\right)}}{4 q_2 \left({q_1+q_2}\right)}
-\frac{q_2\sin^2{\left({2 \phi}\right)}}{4 q_1 \left({q_1+q_2}\right)}
}\right)  
\nonumber\\
&= & -\frac{15\pi^2}{64 \pi r^3_{ll'}}\sin^2{\left({2 \phi}\right)} \  .
\label{EQ:chiAC}
\end{eqnarray}

\medskip
\par

Overall the  dimensionless susceptibility matrix assumes the following compact form:

\begin{eqnarray}
\tilde{\boldsymbol \chi} &=&
\tilde{\boldsymbol\chi}_{\texttt{inter}} \left({r_{ll'}; \phi}\right) = \mathbf{\Phi}\left({\phi}\right) \circ \mathbf{R}_{\texttt{inter}}\left({ r_{ll'}}\right)
\nonumber\\
\mathbf{\Phi} &=&
\mathbf{\Phi}^{1/2} \circ \mathbf{\Phi}^{1/2}
\nonumber\\
\mathbf{R}_{\texttt{inter}} &=& \frac{1}{2 \pi r^3_{ll'}}\frac{\pi^2}{8}
\left({
\begin{matrix}
1  && -3  && -3.7481 \\
{}  && 1 && -3 \\
{}  && {}  && 1 
\end{matrix}
}\right).
\label{BIGU}
\end{eqnarray}
Therefore, although the exchange interaction obeys an inverse  cubic law for the separation between spins located on the $\texttt{A,\ B}$ or $\texttt{C}$ sites, the strength of this coupling is in general determined by the specific lattice involved as well as the hopping parameter $\phi$.   It is only the $\texttt{BB}$ term which is totally independent of $\phi$ since $G_{\texttt{BB}}$ itself does not vary with the hopping strength. Furthermore, we were able to perform the integrals over the closed contour in Fig.\  \ref{fig:2} involving Hankel functions to obtain closed form analytic results for the elements of the matrix $\tilde{\chi} \left({r_{ll'}; \phi}\right)$ in Eq.~(\ref{BIGU}). The $r_{ll'}^{-3}$ law for the exchange interaction was also confirmed in Ref.~\cite{RKKY} along with the fact that the matrix elements are anisotropic. The   advantage  of having the explicit dependence on the coupling parameter $\phi$ in Eq.~(\ref{BIGU}) is that it provides an easy comparison between terms and one could  evaluate their influence in physical phenomena. Interestingly, our results in Eq.~( \ref{BIGU}) show that regardless of the value for $\phi$ ($0<\phi\leq \pi/4$), the diagonal elements are always positive whereas the off-diagonal elements are negative. For the Lieb or dice lattice when $\phi=\pi/4$,  it turns out that the strength of the interaction is largest. These observations confirm that the interaction is ferromagnetic between spins on the same lattice, but antiferromagnetic when they are situated on different lattices.

\subsection{Finite chemical potential and zero magnetic field}

We now turn our attention to setting up a calculation for non-zero chemical potential for the system either by appropriate doping or subjecting the sample to a potential difference with respect to a remote conducting substrate. In order to manage relevant poles contributions here, we focus on $p-$type doping. After an obvious change of the energy variable Eq. \eqref{EQn:1} assumes the form:

\begin{equation}
\label{EQ:1F}
\chi_{\mu \nu}\left({r_{ll'}; \mu_F, \phi}\right)  = -\frac{E_0}{\pi} \oint \limits_{C} d \omega \texttt{Im}\left[{G^2_{\mu\nu}\left({\omega-\mu_F}\right)}\right]  \    ,
\end{equation}
where $\mu_F=E_F/E_0>0$ is the normalized chemical potential. Making use of the definitions of the Green's functions it is straightforward to calculate the susceptibility 
matrix separated as:
\begin{equation}
\tilde{\boldsymbol\chi}  =
\tilde{\boldsymbol \chi}_{\texttt{inter}} \left({r_{ll'}; \phi}\right)
  +
\tilde{\boldsymbol\chi}_{\texttt{intra}} \left({r_{ll'}; \mu_F, \phi}\right)
\end{equation}
In this notation, $\tilde{\boldsymbol\chi}_{\texttt{intra}}$ represents the intra-band contribution to the response.
The interaction between impurities at the $\texttt{A}$ and $\texttt{B}$ sites of the lattice is obtained via residue expansion and the contour integration described above, and is given by

\begin{gather}
\label{EQ:chiABF}
\tilde{\chi}_{\texttt{intra};\texttt{A}\texttt{B}}  =
2 \texttt{Im}\mathcal{H}_1
\left({\frac{ i q_1 \theta\left({\mu_F - q_2}\right) }{q_1^2 - q^2_2}}\right)
\cos^2\phi =\\
\notag=\mu_F^3 \cos^2(\phi)
\mathcal{I}_{\texttt{AB}}\left({\mu_F r_{ll'}}\right) \; ;\\
\notag
\mathcal{I}_{\texttt{AB}}\left({x}\right)
= \frac{1}{x^3}\texttt{Im} \int \limits_0^{x}
dq \  q^2 J_1\left({q}\right) K_1\left({-i q}\right).
\end{gather}

\begin{figure}[h!]
\centering
\includegraphics[width=0.35\textwidth]{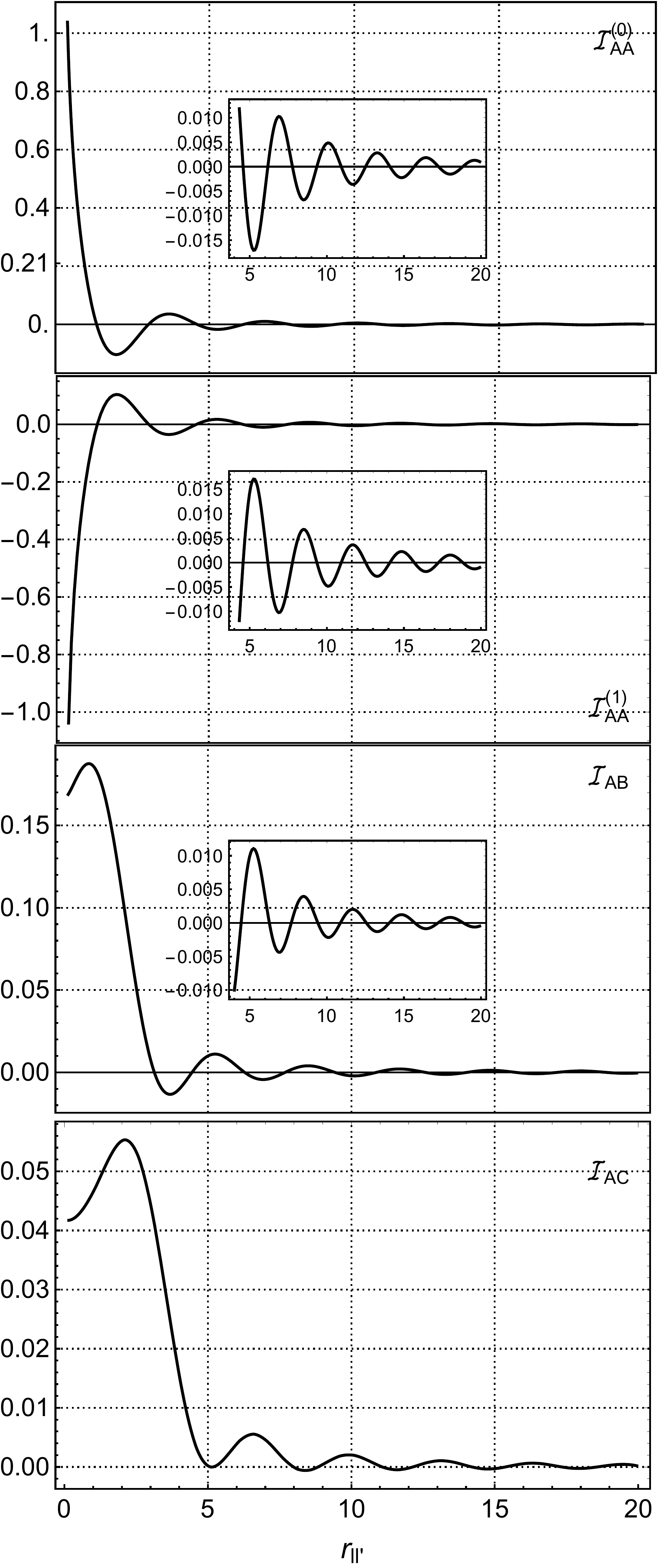}
\caption{Set of integrals involved in the intraband contributions to the susceptibility as functions of the spin separation $r_{ll'}$. For  clarity, the insets display the results over a smaller range of separation between the impurities.}
\label{fig:3}
\end{figure}

The pre-factor of $2$ takes account of the symmetry of the problem upon  the interchange of the variables $q_1 \leftrightarrow q_2$. The symmetric positions of the poles also assure us  that changing to $n-$doping results in $\tilde{\chi}_{\texttt{intra},\texttt{A}\texttt{B}} \rightarrow - \tilde{\chi}_{\texttt{intra},\texttt{A}\texttt{B}}$.
In similar fashion we obtain the  contribution to the susceptibility between the $\texttt{A}$ and  $\texttt{C}$ sub-lattices
\begin{gather}
\tilde{\chi}_{\texttt{intra},\texttt{A}\texttt{C}} =
2 \texttt{Im}\mathcal{H}_2
\left({-\frac{ i q_2^3 \theta\left({\mu_F - q_1}\right) }
{4 q_1 q_2 \left({q_1^2 - q^2_2}\right)}}\right)
\sin^2\left({2 \phi}\right)=\\
\notag
=\mu_F^3 \sin^2\left({2 \phi}\right)
\mathcal{I}_{\texttt{AC}}\left({\mu_F r_{ll'}}\right) \; ;\\
\notag
\mathcal{I}_{\texttt{AC}}\left({x}\right)
= -\frac{4 \pi}{x^3} \int \limits_0^{x}
dq \ q^{-2} J_2\left({q}\right)
\mathcal{G}_{3,1}^{0,2}
\left({\frac{4}{q^2} \vert
\begin{smallmatrix}
-2,& 0, & 1/2\\
1/2
\end{smallmatrix}
}\right),
\end{gather}
with $\mathcal{G}$ standing for the Meijer G function. Finally we obtain the $\texttt{A}$, $ \texttt{A}$ term as
\begin{gather*}
\tilde{\chi}_{\texttt{intra},\texttt{A}\texttt{A}}  
\\
\notag
=2 \texttt{Im}\mathcal{H}_0
\left({
-i\frac{ q_1^3 - q_1 q_2^2 \sin^2\phi }
{q_1^2 - q^2_2}
\theta\left({\mu_F - q_1}\right)
}\right)
\cos^2 \phi \\
\notag
= \mu_F^5
\cos^2 \phi
\left[{
\mathcal{I}_{\texttt{AA}}^{\left({0}\right)}\left({\mu_F r_{ll'}}\right)
+
\mathcal{I}_{\texttt{AA}}^{\left({1}\right)}\left({\mu_F r_{ll'}}\right)
\sin^2\phi}\right] \; ;\\
\notag
\mathcal{I}_{\texttt{AA}}^{\left({0}\right)}\left({x}\right)=
\frac{1}{x^5} \int \limits_{0}^{x}
dq \  q^4 J_0 \left({q}\right) 2 K_0\left({-i q}\right) \; ,\\
\notag
\mathcal{I}_{\texttt{AA}}^{\left({1}\right)}\left({x}\right)=
\frac{1}{x^5} \int \limits_{0}^{x}
dq \  q^4 J_0 \left({q}\right)
\left[{ J_0\left({q}\right)\log\left({q^2/4}\right) + }\right.\\
\left.{+2 \mathcal{H}^{\left({1,0}\right)} \left({1,-q^2/4}\right)}\right]\ .
\end{gather*}
Here,  $\mathcal{H}^{\left({1,0}\right)}$ is the regularized confluent hypergeometric function. Note that the numerical simulations (see Fig.~\ref{fig:3}) reveal $\mathcal{I}_{\texttt{AA}}^{\left({1}\right)}\left({x}\right)=-\mathcal{I}_{\texttt{AA}}^{\left({0}\right)}\left({x}\right)$. Therefore, we can rewrite the intraband contribution in a form similar to the interband result, i.e.,
\begin{equation}
  \tilde{\chi}_{\texttt{intra},\texttt{A}\texttt{A}} \left({r_{ll'}; \mu, \phi}\right)=
  \mu_F^5
\cos^4 \left({\phi}\right) \mathcal{I}_{\texttt{AA}}\left({\mu_F r_{ll'}}\right)\ ,
\end{equation}
where $\mathcal{I}_{\texttt{AA}} = \mathcal{I}_{\texttt{AA}}^{\left({0}\right)}$.
Finally, we obtain the intraband susceptibility matrix given as

\begin{gather}
\tilde{\boldsymbol\chi}_{\texttt{intra}} \left({r_{ll'}; \mu_F, \phi}\right) = \mathbf{\Phi}\left({\phi}\right) \circ \mathbf{R}_{\texttt{intra}}\left({\mu_F, r_{ll'}}\right)
\\
\notag
\mathbf{R}_{\texttt{intra}} =\mu_F^3
\left({
\begin{matrix}
\mu_F^2 \mathcal{I}_{\texttt{AA}}  && \mathcal{I}_{\texttt{AB}}  && \mathcal{I}_{\texttt{AC}} \\
\mathcal{I}_{\texttt{AB}} && \mu_F^2 \mathcal{I}_{\texttt{AA}} && \mathcal{I}_{\texttt{AB}} \\
\mathcal{I}_{\texttt{AC}}  && \mathcal{I}_{\texttt{AB}}  && \mu^2_F\mathcal{I}_{\texttt{AA}} 
\end{matrix}
}\right)
\end{gather}
 The graphs in Fig.~\ref{fig:3}  show that when the system has finite chemical potential, the exchange interaction can vary between ferromagnetic and antiferromagnetic as the spin separation is increased. This is true regardless of the chosen value for the parameter $\phi$.

\subsection{Magnetic field effects on the RKKY interaction}
\label{sec4}

 We shall perform our calculations using the Landau gauge, for which the vector potential is ${\bf A}=-B_z y\hat{x}$  and $\nabla \times {\bf A}= B_z\hat{z}$ is the magnetic field. Using that Hamiltonian in Eq.\ (\ref{eqn:1}),  one can determine the wave functions and Landau levels for the lattice. Making use of the vector potential ${\bf A}=-B_z y\hat{x}$ and the Peierls substitution $\hbar{\bf k}\to {\bf p}\to {\bf p}+e{\bf A}$,  where $\hbar{\bf k}$  is the momentum  eigenvalue in the absence of magnetic field and ${\bf p}$ is the momentum operator,  we have

\begin{gather}
\label{Hamil}
\mathbf{\hat{H}}_K=-\mathbf{\hat{H}}_{K'}^\ast=\\
\notag=E_B \begin{pmatrix}&&0 &&\cos\phi \ {\hat a} &&0\\
&&\cos\phi\ {\hat a}^+&& 0 &&\sin\phi\  {\hat a} \\ &&0 &&\sin\phi\ {\hat a^+}&& 0\end{pmatrix}  \  ,
\end{gather}
where $E_B = \sqrt{2} \gamma l_H^{-1}$ is the cyclotron energy related to the magnetic length $l_H = \sqrt{\hbar/\left({e B_z}\right)}$. 
We also define the destruction operator ${\hat a}=\frac{1}{\sqrt{2\hbar e B_z}}({\hat p_x}-e B_z {\hat y}-i {\hat p_y})$ and the creation operator ${\hat a^+}=\frac{1}{\sqrt{2\hbar e B_z}}({\hat p_x}-e B_z {\hat y}+i {\hat p_y})$ as  for the harmonic oscillator.   We note that when $\phi=0$, the Hamiltonian sub-matrix consisting of the first two rows and columns is exactly that used in \cite{GG1,Roldan} for mono-layer graphene.
\medskip
\par
In the most general case, let us denote the eigenstates by $\{\mathbf{\Psi}_{\mathbf{n}}\left({\mathbf{r}}\right),E_{\mathbf{n}}\}$, where the eigenfunctions are orthonormal,i.e., $\int d^2 \mathbf{r} \mathbf{\Psi}^{T}_{\mathbf{n1}}\left({\mathbf{r}}\right) \mathbf{\Psi}^{\star}_{\mathbf{n2}}\left({\mathbf{r}}\right) = \delta_{\mathbf{n1},\mathbf{n2}}$.  We then write the Green's function as

\begin{equation}
    \mathbf{G}\left({E;\mathbf{r}_{ll'}}\right) = \frac{1}{E\mathbf{I}-\mathbf{H}}
    = \sum \limits_{\mathbf{n}} \frac{\mathbf{\Psi}^{\star}_{\mathbf{n}}\left({\mathbf{r}_{l}}\right) \mathbf{\Psi}^{T}_{\mathbf{n}}\left({\mathbf{r}_{l'}}\right)}{E-E_{\mathbf{n}}+i 0^{+}} \  ,
    \label{EQ:Greensfunctionexpansion}
\end{equation} 
In the presence of magnetic field, we have $\mathbf{n} = \left\{\lambda,s,n,k_y\right\}$, where $\lambda = \pm 1$ denotes the valley for $\mathbf{K}$ and $\mathbf{K'} = - \mathbf{K}$ respectively; $s = -1,0,1$ stands for the valence, flat and conduction bands respectively; $n \geq 0$ is the Landau level index; and $k_y$ is the wave vector. The energies are given by diagonalizing Hamiltonian \eqref{Hamil}:
\begin{equation}
    \label{EQ:Landauenergies}
    E_{\mathbf{n}} = E_B\epsilon_{\lambda,s,n} = E_B s \sqrt{n+\chi_\lambda}  \ ,
\end{equation}
Here the auxiliary parameter $\chi_{\lambda} = \left[{1-\lambda \cos \left({2 \phi}\right)}\right]/2$ has been used where $0 \leq \chi_{\lambda} <1$.
\par
The susceptibility components at $T=0$ K and Fermi energy $E_F$ are given by Eqs.~\eqref{EQ:1F} and \eqref{EQ:Greensfunctionexpansion} as

\begin{gather}
\label{EQ:suceptibility}
    \chi_{\mu\nu}   = -\frac{1}{\pi}\texttt{Im} \int \limits_{-\infty}^{\infty} dE\  \theta\left({E_F-E}\right)
    G^2_{\mu\nu}\left({E;  \mathbf{r}_{ll'}}\right)  =  \\
    \notag=
    -\frac{1}{\pi} \texttt{Im} \sum \limits_{\mathbf{n}1,\mathbf{n}2} \Psi^{\mu\nu}_{\mathbf{n}1;\mathbf{n2}} \left({\mathbf{r}_{l},\mathbf{r}_{l'}}\right) 
   \int \limits_{-\infty}^{\infty} dE\   \frac{\theta\left({E_F-E}\right)}{\left({E_{\mathbf{n}1}-E_{\mathbf{n}2}}\right)}
   \times\\
   \notag
   \times
   \left({
   \frac{1}{E-E_{\mathbf{n}1}+i 0^{+}}-
   \frac{1}{E-E_{\mathbf{n}2}+i 0^{+}}
   }\right) = \\
   \notag
   =
    \sum \limits_{\mathbf{n}1,\mathbf{n}2} \Psi^{\mu\nu}_{\mathbf{n}1;\mathbf{n2}} \left({\mathbf{r}_{l},\mathbf{r}_{l'}}\right)
    \left[{
    \frac{\theta\left({E_F-E_{\mathbf{n}1}}\right) - \theta\left({E_F-E_{\mathbf{n}2}}\right)}{E_{\mathbf{n}1}-E_{\mathbf{n}2}}
    }\right] \  .
\end{gather}
Here we have used the shorthand notation $\Psi^{\mu\nu}_{\mathbf{n}1;\mathbf{n2}} \left({\mathbf{r}_{l},\mathbf{r}_{l'}}\right) = \Psi^{\star\mu}_{\mathbf{n}1}\left({\mathbf{r}_{l}}\right)\Psi^{\nu}_{\mathbf{n}1}\left({\mathbf{r}_{l'}}\right)
\Psi^{\star\mu}_{\mathbf{n}2}\left({\mathbf{r}_{l'}}\right)\Psi^{\nu}_{\mathbf{n}2}\left({\mathbf{r}_{l}}\right) $.
\par
Mapping the sites labels $\texttt{A,B,C} \rightarrow -1,0,1$ and separating the spacial variables in the wave function we obtain
\begin{equation}\label{Eq:Psi}
  \Psi^{\star\mu}_{\mathbf{n}}\left({\mathbf{r}_{l}}\right) = \psi ^\mu_{\lambda,s,n} \phi_{n+\lambda \mu,k_y}\left({x_l}\right) \texttt{e}^{-i k_y y_{l}} \texttt{e}^{-i \lambda K_y y_{l}},
\end{equation}
where the vector components specific to the given lattice are denoted by $\psi ^\mu_{\lambda,s,n}$ and  $\phi_{n,k_y}\left({x_l}\right)$ are the harmonic oscillator wave functions.
When $s^2=1$ these components assume the following form
\begin{equation}\label{Eq:psi}
    \psi^{\mu}_{\lambda,s,n}=\frac{1}{\sqrt{2 \left({ n+\chi_{\lambda}}\right)}}
    \begin{cases}
    \sqrt{n\left({1-\chi_{\lambda}}\right)}, & \lambda\mu=-1\\
    s\lambda\sqrt{(n+\chi_{\lambda})}, & \lambda\mu=0 \ .\\
    \sqrt{(n+1)\chi_\lambda}, & \lambda\mu=1\\
    \end{cases}
\end{equation}
For the flat band ($s=0$) when $n>0$ the components are
\begin{equation}
    \psi^{\mu}_{\lambda,s,n}=\frac{1}{\sqrt{{ n+\chi_{\lambda}}}}
    \begin{cases}
    -\lambda\sqrt{(n+1)\chi_{\lambda}}, & \lambda\mu=-1\\
    0, & \lambda\mu=0 \ ,\\
    \lambda\sqrt{n(1-\chi_\lambda)}, & \lambda\mu=1\\
    \end{cases}
\end{equation}
while for $n=0$ the components are
\begin{equation}
    \psi^{\mu}_{\lambda,s,n}=
    \begin{cases}
    0, & \lambda\mu=-1\\
    0, & \lambda\mu=0 \ . \\
    1, & \lambda\mu=1\\
    \end{cases}
\end{equation}

By combining Eqs. \eqref{EQ:Landauenergies}, \eqref{EQ:suceptibility} and \eqref{Eq:psi}, after some algebra (see Appendix \ref{AP:3}) we finally obtain the general form of the susceptibility components:
\begin{gather}
\label{EQ:main}
 \chi^{\mu\nu}  =
    \frac{\mathcal{A}}{E_B \left({2 \pi l_H}\right)^2} \tilde{\chi}^{\mu\nu}
\left({ \mathbf{r}_l,\mathbf{r}_{l'}}\right),\\
    \notag
\tilde{\chi}^{\mu\nu}\left({ \mathbf{r}_l,\mathbf{r}_{l'}}\right) =
    \sum \limits_{\lambda_{1,2}=\pm 1} \sum \limits_{s_{1,2}=0,\pm 1} \sum \limits_{n_{1,2}\geq 0}
    \psi^{\mu\nu}_{\lambda_{1}s_{1}n_{1};\lambda_{1}s_{1}n_{1}} \times
    \\
    \notag
   \tilde{\Phi}_{n_1+\lambda_1 \mu}^{n_1+\lambda_1 \nu} \left({s_1;\mathbf{r}_l,\mathbf{r}_{l'}}\right)
    \tilde{\Phi}_{n_2+\lambda_2 \mu}^{n_2+\lambda_2 \nu} \left({s_2;\mathbf{r}_{l'},\mathbf{r}_{l}}\right)
    \texttt{e}^{-i K \left({\lambda_1-\lambda_2}\right) \left({y_l-y_{l'}}\right)}  \\
    \notag
    \times
    \frac{\theta\left({\mu_F-s_1 \sqrt{n_1+\chi_{\lambda_1}}}\right) - \theta\left({\mu_F-s_2 \sqrt{n_2+\chi_{\lambda_2}}}\right)}{s_1 \sqrt{n_1+\chi_{\lambda_1}} - s_2 \sqrt{n_2+\chi_{\lambda_2}}},
\end{gather}
where we have introduced the normalized Fermi energy $\mu_{F} = E_F/E_B$ as well as $\psi^{\mu\nu}_{\lambda_{1}s_{1}n_{1};\lambda_{1}s_{1}n_{1}}= \psi_{\lambda_{1},s_{1},n_1}^{\mu} \psi_{\lambda_{1},s_{1},n_1}^{\nu} \psi_{\lambda_{2},s_{2},n_2}^{\mu}\psi_{\lambda_{2},s_{2},n_2}^{\nu}$. Equation\  \eqref{EQ:main} is applicable for a wide range of experimental parameters and serves as a basis for numerical simulations which are presented below. For simplicity, we neglect highly oscillatory inter-valley terms setting $\lambda_1=\lambda_2 =\lambda=\pm 1$.

\medskip
\par
\begin{figure}[h!]
\centering
\includegraphics[width=0.45\textwidth]{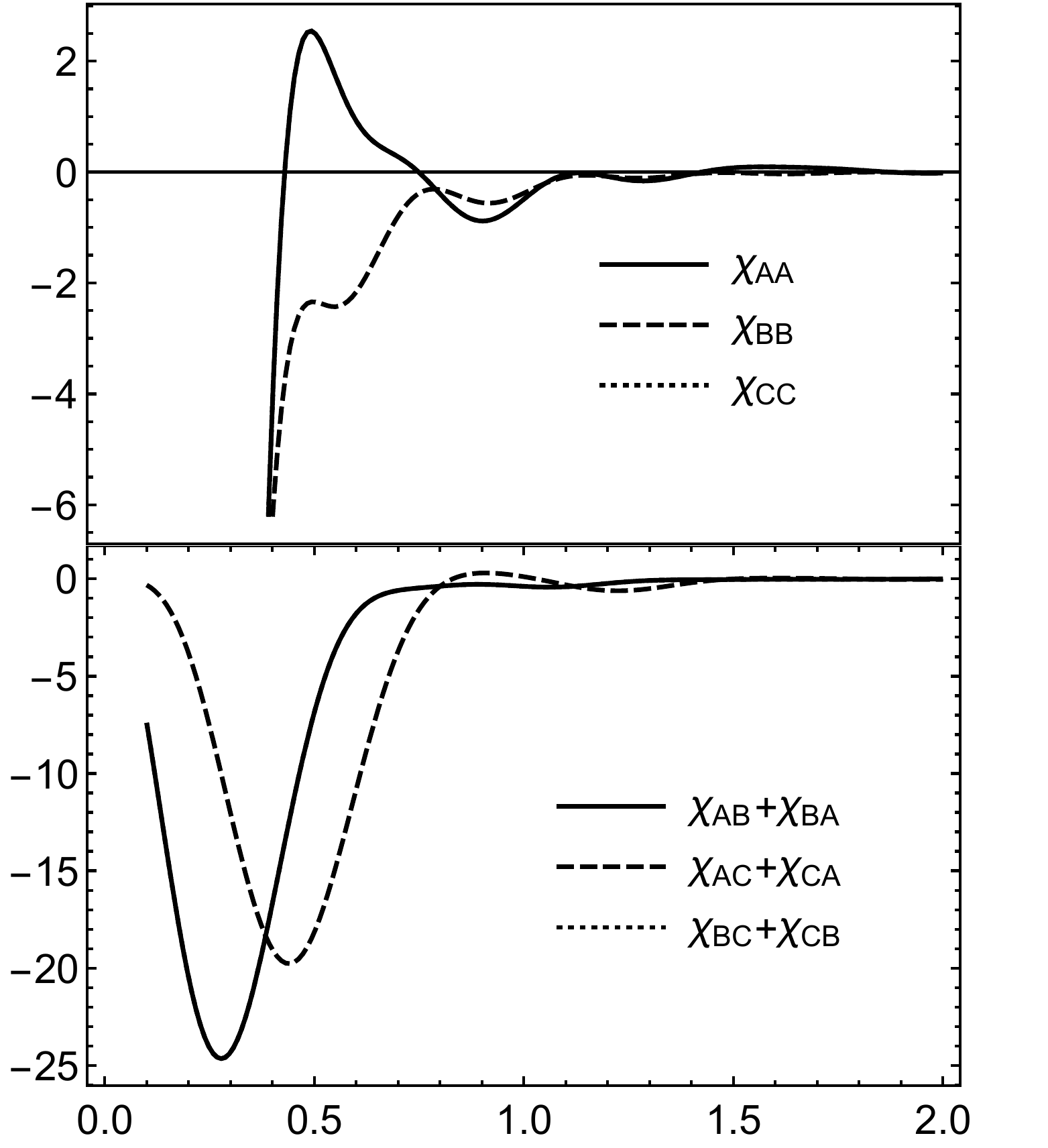}
\caption{Spin susceptibility as a function of the inter-particle separation for $E_F=0\ ,\ T=0$\ K, $\phi=\pi/4$.}
\label{Fig:4}
\end{figure}
\begin{figure}[h!]
\centering
\includegraphics[width=0.45\textwidth]{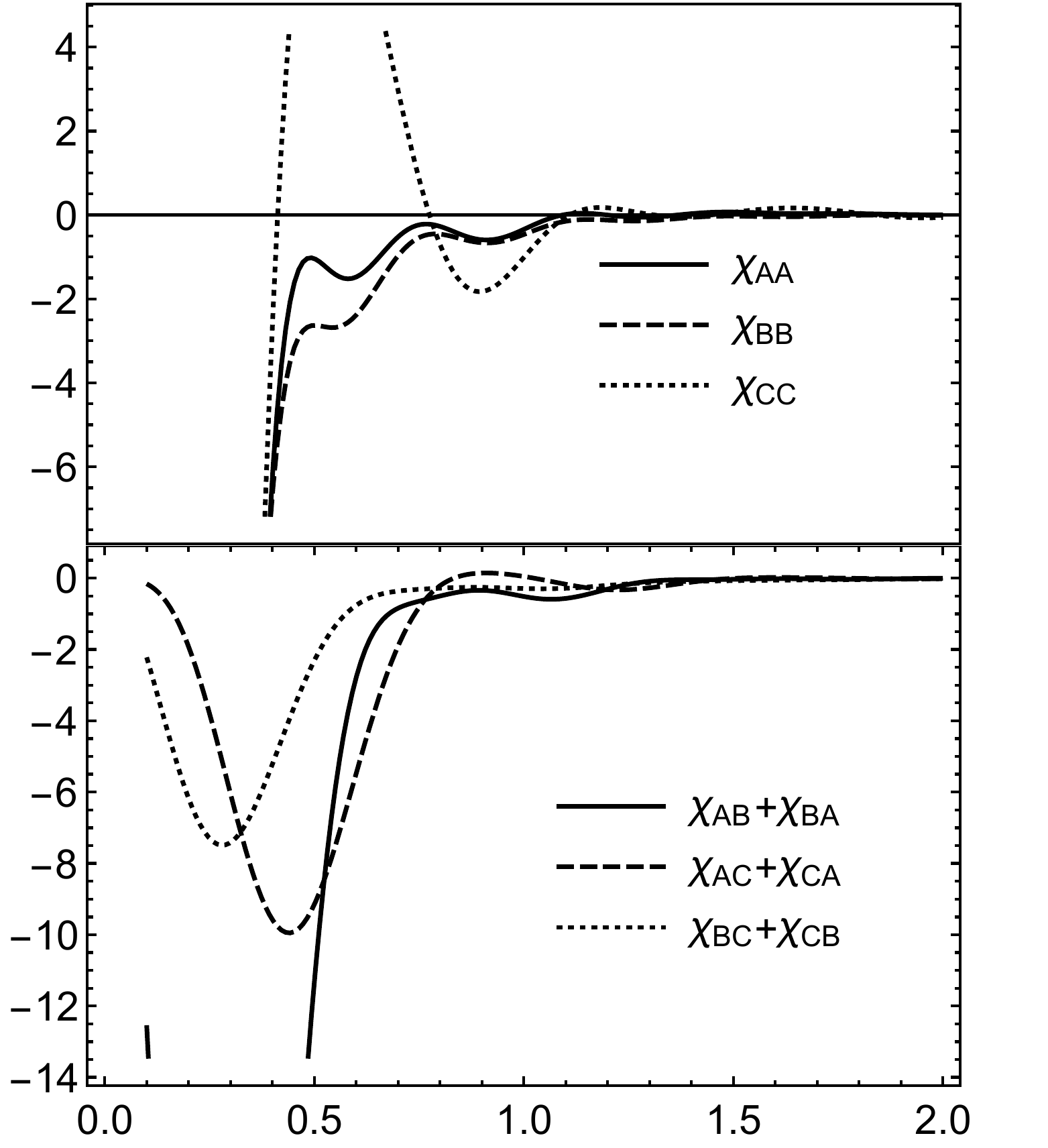}
\caption{Spin susceptibility as a function of the inter-particle separation for $E_F=0,\ T=0$\ K, $\phi=\pi/8$.}
\label{Fig:5}
\end{figure}
\begin{figure}[h!]
\centering
\includegraphics[width=0.45\textwidth]{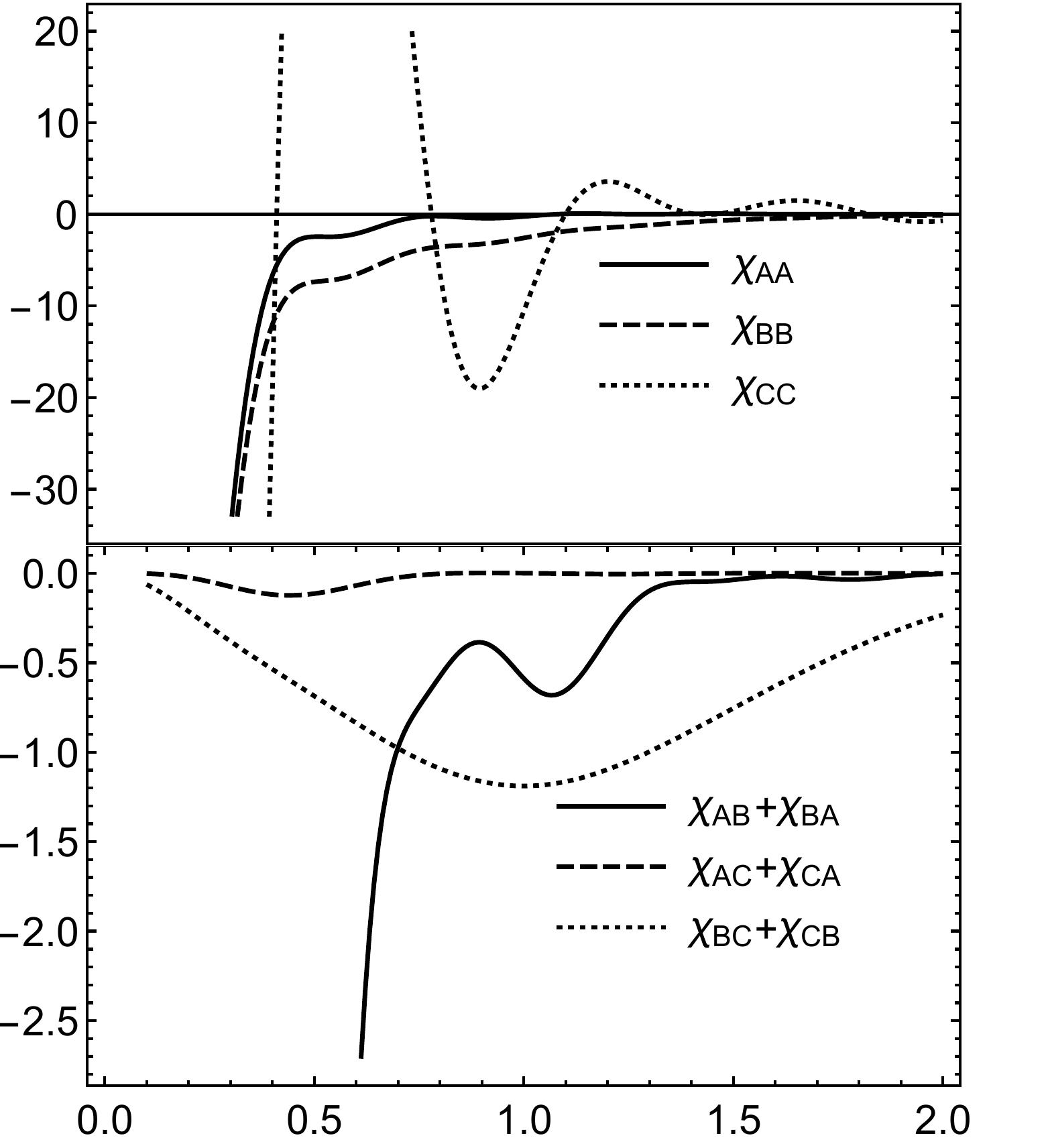}
\caption{Spin susceptibility as a function of the inter-particle separation for  $E_F=0,\ T=0$\ K, $\phi=\pi/80$.}
\label{fig:6}
\end{figure}

Figures\  \ref{Fig:4},  \ref{Fig:5}  and  \ref{fig:6} present the magnetic field dependent susceptibility as a function of the spin separation.  The structure has $E_F=0$  at $T=0$ K.  
Three values of $\alpha$ were chosen in the numerical calculations  All chosen $\phi $  show regions of ferromagnetic and antiferromagnetic behavior with the amplitude of the oscillations decreasing with increasing separation between the spins on the lattice. 
However, for $\phi=\pi/80$ in Fig.~\ref{fig:6}, 
$\chi_{\texttt{CC}}$ has the largest amplitude oscillations and 
$\chi_{\texttt{AB}}  +  \chi_{\texttt{BA}}$,  $\chi_{\texttt{AC}}  +  \chi_{\texttt{CA}}$
and $\chi_{\texttt{BC}}  +  \chi_{\texttt{BA}}$ all remain negative independent of $r_{ll'}$.   
These results are interesting as they demonstrate how one could control the magnetic behavior of $\alpha$-${\cal T}_3$.

\medskip
\par

\begin{figure}[h!]
\centering
\includegraphics[width=0.45\textwidth]{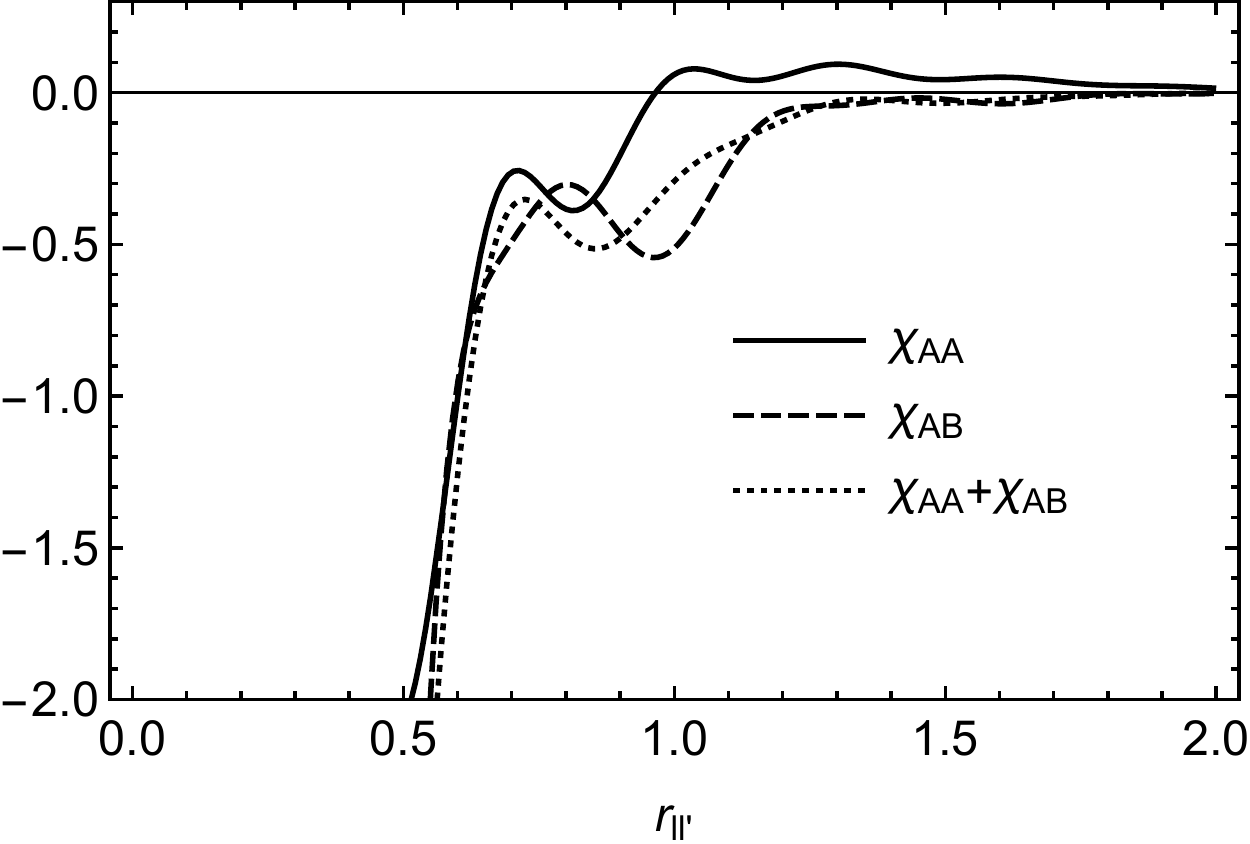}
\caption{Spin susceptibility as a function of the inter-particle separation for graphene with  $E_F=0\ ,\ T=0$ K.}
\label{Fig:7}
\end{figure}

Most importantly, these results in Figs.\  \ref{Fig:4},  \ref{Fig:5}  and  \ref{fig:6} signal that the magnetic properties of the $\alpha$-${\cal T}_3$ lattice near $\alpha=0$ need to be compared with those for graphene in Fig.\  \ref{Fig:7}. Remarkably, the susceptibility has one sign for small $r_{ll'}$.  The component $\chi_{\texttt{AA}}$  oscillates but remains positive for large spin separation.  On the contrary, both $\chi_{\texttt{AB}}$ and the sum $\chi_{\texttt{AA}}+\chi_{\texttt{AB}}$ remain negative in this limit.  This behavior is independent of the position of the Fermi level.  We note that in doing the calculations for graphene, we {\em first} set $\alpha=0$ in Eq.~(\ref{Hamil}) before calculating the eigenstates which were in turn  employed in the spin susceptibility. Therefore, the change in behavior discovered here is clear when $\alpha$ is finite and zero.

\section{\label{sec4}Limiting cases in magnetic field}

We now turn our attention to two specific cases where closed form analytic expressions can be obtained for the spin susceptibility.  A very intriguing case occurs in strong magnetic field for which there are well separated Landau levels at $\lambda = 1$ and $\phi \to 0$. Assuming an undoped lattice configuration, i.e. $\mu_F=0$, the dominant  terms come from $n_{1,2} = 0$ contributions to Eq.\ \eqref{EQ:main}.   We have 
 
\begin{gather}
\tilde{\chi}^{\mu\nu} =  
\sum \limits_{s_{1,2}=0,\pm 1}
   \tilde{\Phi}_{\mu}^{\nu} \left({s_1;\mathbf{r}_l,\mathbf{r}_{l'}}\right)
   \tilde{\Phi}_{\mu}^{\nu} \left({s_2;\mathbf{r}_{l'},\mathbf{r}_{l}}\right)  \times \\
   \notag
   \psi_{1,s_{1},0}^{\mu} \psi_{1,s_{1},0}^{\nu} \psi_{1,s_{2},0}^{\mu}\psi_{1,s_{2},0}^{\nu}
   \frac{\theta\left({-s_1 \sin \phi}\right) - \theta\left({-s_2 \sin \phi}\right)}{s_1 \sin \phi - s_2 \sin \phi} \  . 
\end{gather}
Let us introduce the normalized temperature $\tilde T = \frac{k_B T}{E_B}$ and the integral representation of the heat kernel instead of $\theta$ function. For  an arbitrarily chosen small temperature, we set $\tilde T = \sin ^2(\phi)$, and expanding the above equation around small positive $\phi$ we obtain:

\begin{gather}
\tilde{\chi}^{\mu\nu} \sim
    \frac{\texttt{Erf}\left[{\frac{1}{\sqrt{2}}}\right]\texttt{Exp}\left[{\frac{-r_{ll'}^2}{2}}\right]}{4 \phi} \times\\
    \notag
   \left[{
   \left({
   \begin{matrix}
   0 & 0 & 0 \\
   0 & -1 & 1 \\
   0 & 1 & -1
   \end{matrix}
   }\right) +
    \left({
   \begin{matrix}
   0 & 0 & 0 \\
   0 & 0 & 0 \\
   0 & 0 & -4
   \end{matrix}
   }\right)
   }\right] \  .
\end{gather} 
The first matrix is due to  transitions between the valence and conduction bands as well as within the conduction band from below to above the Fermi level.   The second matrix arises from transitions from the flat band to the conduction band.  The upshot from these results is that the largest change in the spin susceptibility occurs in the limit when $\phi\to 0$ and there is no smooth transition from finite $\phi$ to $\phi=0$, thereby indicating that there is a phase transition between graphene ($\phi=0$)  and the $\alpha$-${\cal T}_3$ model. This anomaly is short range due to the exponent, and has no counterpart in the $\lambda = -1$ valley. 

\medskip
\par
We also study the case of weak magnetic field $E_B$ or high doping $E_F$, so that the Fermi level $n_F$ is defined via $\sqrt{n_F - 1 + \chi_{\lambda_1}}\leq  \mu_F \leq \sqrt{n_F  + \chi_{\lambda_2}}$. There are only intra-band $s_{1}=s_2=1$ contributions. The leading terms (largest contributions to the sum) come from the states nearest to $n_F$. Specifically, for large $n_F$, we found numerically that the terms in Eq.\  \eqref{EQ:main} scale as  $\delta_{|n_1-n_2|,1}$. The transitions from the flat to the conduction band do not follow this rule, the rather scale as $\sim  1/n_F$ which allows us to neglect such contributions. A similar approach was adapted by Lozovik \cite{Lozovik} when he discussed edge magnetoplasmons in graphene (leading contributions to the conductivity tensor in the aforementioned limit).  However, there is an important difference in that the magnetoplasmons are given by the optical conductivity tensor where $\delta_{|n_1-n_2|,1}$ is the  true selection rule which applies for all $n$. 
\par
In this limiting case Eq.\eqref{EQ:main} can be written in a compact form:
\begin{gather*}
   \tilde{\boldsymbol \chi} = \left[{\mathbf{I} \circ \mathbf{\Phi} + \mathbf{V}_{\lambda_{1}=-\lambda_{2}} \circ \mathbf{\Phi}_{\lambda_{1}=-\lambda_{2}} }\right] \circ \mathbf {R} \ .
\end{gather*}
Contributions from the same valley $\lambda_1=\lambda_2$  (first term in the square brackets of the above expression) are given by  $\mathbf{\Phi}\left({\phi}\right)$ which is identical to the no-magnetic field case Eq.\  \eqref{BIGU}.  However, for mixed valley contributions, $\lambda_1=-\lambda_2$, we obtain highly oscillatory terms $\mathbf{V}_{\lambda_{1}=-\lambda_{2}} = \cos (2 K y_{ll'}) \mathbf{I}$ along with a peculiar form for the $\phi-$matrix:
 
 \begin{equation}
 \label{EQ:phi2}
    \mathbf{\Phi}_{\lambda_{1}=-\lambda_{2}}(\phi)= \left(
\begin{array}{ccc}
 \frac{1}{4} \cot ^2\phi  & \frac{1}{2}  \csc ^2\phi  & -2  \\
 \frac{1}{2}  \csc ^2\phi  &  \csc ^2(2 \phi ) & \frac{1}{2}  \sec ^2\phi  \\
 -2  & \frac{1}{2} \sec ^2\phi  & \frac{1}{4} \tan ^2\phi  \\
\end{array}
\right)   \  .
 \end{equation}
It is informative to look at the upper-left $2 \times 2$ sub-matrix in Eqs.\eqref{BIGU} and \eqref{EQ:phi2} corresponding to the graphene like case of $\texttt{A}$ and $\texttt{B}$ sub-lattices. While Eq.\eqref{BIGU} provides smooth transition to graphene at $\phi \rightarrow 0$, the valley mixing in Eq.\eqref{EQ:phi2} gives $\phi^{-2}$ scaling. The absence of the smooth graphene limit can be directly attributed to broken symmetry for $K$ and $K'$ valleys in magnetic field.

\medskip
\par
The site-to site distance and Fermi number dependent matrix referred to above is  given by
\begin{widetext}
\begin{equation}
\label{EQ:R}
 \mathbf R(r_{ll'},n_F)=\frac{1}{2\pi r}\left(
\begin{array}{ccc}
 -4 \cos ^2\left(2 \sqrt{n_F} r\right) & e^{-r^2} \cos \left(4 \sqrt{n_F} r\right)+1 & \frac{1}{4} \left[e^{-r^2} \cos \left(4 \sqrt{n_F} r\right)+1\right] \\
 {} & -4 \cos ^2\left(2 \sqrt{n_F} r\right) & e^{-r^2} \cos \left(4 \sqrt{n_F} r\right)+1 \\
 {} & {} & -4 \cos ^2\left(2 \sqrt{n_F} r\right) \\
\end{array}
\right)  \   , 
\end{equation}
\end{widetext}
where for convenience we have omitted the subscripts  $r_{ll'}/\sqrt{2} \to r$.  If we formally associate $\sqrt{n_F}$ with $k_F$, the oscillations in the above equation correspond to Kohn anomalies in the absence of magnetic field which was first reported in Ref.\  \cite{O8}. However, they are much larger in range due to the $\sim 1/r$ dependence.  At larger distances, we can neglect the terms $\sim \exp({-r^2})$ and the oscillations for impurities which are placed on different sub-lattices vanish.

\newpage

\newpage

\section{ Concluding Remarks and Summary}
\label{sec5}

We have investigated the  behavior of the RKKY interaction for  undoped and doped $\alpha$-${\cal T}_3$ semi-metals as well as when they are subjected to a uniform perpendicular magnetic field. Specifically, we have shown the following: (a) For undoped samples, the RKKY interaction obeys an inverse cubic law for the separation between spins located on lattice sites. The strength of this interaction is anisotropic and determined by the adjustable hopping parameter $\phi$ except when both spins are on   $\texttt{B}$ sites. Furthermore the $\texttt{AA}$, $\texttt{BB}$ and $\texttt{CC}$ exchange interactions are ferromagnetic but the sign of this interaction is reversed when the spins are located on different sub-lattices; (b) for the case when the chemical potential is finite, we were able to express  our closed form analytic expression for the spin susceptibility in the same algebraic form as the case (a). However, the amplitudes of these interactions are multiplied by an oscillatory factor which could be positive or negative for ranges of the spin separations; (c) in the presence of magnetic field, the spin susceptibility oscillates as the spin separation is varied displaying ranges of ferromagnetism and antiferromagnetism. When $\phi$ is small, we found that the behavior of the susceptibility is radically different compared to when the dice or Lieb phase  ($\phi=\pi/4$) is approached. These observations confirm that a phase transition occurs as $\phi\to 0$ and this phase change is signaled  through an applied magnetic field; (d)    we were able to obtain analytic expressions for the spin susceptibility in the limit of low magnetic field or high doping.  Interestingly, the power law behavior as a function of spin separation is $\sim 1/r$ which is a new result reported here. At large distances between the impurities RKKY interaction exhibits Kohn anomalies  only when those are located on the same sub-lattices. These effects are experimentally observable signatures of the  electronic properties of $\alpha$-${\cal T}_3$ semi-metals and could serve to motivate others  to apply them to  future technologies.
  
\appendix

\section{\label{AP:1} Derivation of Eq.~(\ref{eqn:GG1})}

The eigenfunction decomposition of the Hamiltonian yields the following representation of the Greens tensor components

\begin{equation*}
G_{\mu\nu}({\bf k},E;\lambda;\phi) =\sum_{s=0,\pm 1} \frac{\Psi^\mu_{s,\lambda,\mathbf{k}} \Psi^{\nu\ast}_{s,\lambda,\mathbf{k}}}{E-s\epsilon_k}\ .
\label{sugg1}
\end{equation*}
Let us focus for example on the first raw in Eq.~\eqref{eqn:GG1},

\begin{eqnarray*}
G_{\texttt{AA}} &=&   \frac{|\Psi^{\texttt{A}}_{1,\lambda,\mathbf{k}} |^2 }{E-\epsilon_k}  +
\frac{|\Psi^{\texttt{A}}_{-1,\lambda,\mathbf{k}} |^2 }{E+\epsilon_k} +
\frac{|\Psi^{\texttt{A}}_{0,\lambda,\mathbf{k}} |^2 }{E }
\nonumber\\
&=&\frac{1}{2} \frac{\cos^2\phi}{E-\epsilon_k} +
\frac{1}{2} \frac{\cos^2\phi}{E+\epsilon_k} +
 \frac{\sin^2\phi}{E }\ ,
\label{sugg2}
\end{eqnarray*}
which clearly adds up to give the first element in that equation. Similarly we obtain the other two components

\begin{eqnarray*}
G_{\texttt{AB}} &=&    \frac{ \Psi^{\texttt{A}}_{1,\lambda,\mathbf{k}} \Psi^{\texttt{B} \ast}_{1,\lambda,\mathbf{k}} }{E-\epsilon_k} +
\frac{ \Psi^{\texttt{A}}_{-1,\lambda,\mathbf{k}} \Psi^{\texttt{B} \ast}_{-1,\lambda,\mathbf{k}}}{E+\epsilon_k}
\nonumber\\
&=& \left( \frac{\lambda}{2} \right)  \frac{e^{-i \lambda\theta_k}\cos\phi}{E-\epsilon_k}   -
\left(\frac{\lambda}{2}  \right) \frac{e^{-i \lambda\theta_k}\cos\phi}{E+\epsilon_k} \ ,
\label{sugg3}
\end{eqnarray*}
and

\begin{eqnarray*}
G_{\texttt{AC}} &=&    \frac{ \Psi^{\texttt{A}}_{1,\lambda,\mathbf{k}} \Psi^{\texttt{C} \ast}_{1,\lambda,\mathbf{k}} }{E-\epsilon_k} +
\frac{\Psi^{\texttt{A}}_{-1,\lambda,\mathbf{k}} \Psi^{\texttt{C} \ast}_{-1,\lambda,\mathbf{k}} }{E+\epsilon_k}
+\frac{ \Psi^{\texttt{A}}_{0,\lambda,\mathbf{k}} \Psi^{\texttt{C} \ast}_{0,\lambda,\mathbf{k}} }{E }
\nonumber\\
&=& \left[ \frac{f_{\lambda,\mathbf{k}}}{\epsilon_k} \right]^2 \cos\phi \sin\phi \left(  \frac{E}{E^2-\epsilon_k^2}   -  \frac{1}{E}  \right) \ ,
\label{sugg7}
\end{eqnarray*}
which again agrees with the result of the direct Green's tensor diagonalization.  This alternative definition of the Green's function is particularly useful when considering magnetic field effects.

\section{\label{AP:2} Derivation of Eq.~(\ref{EQn:GreensFunction})}
Here we obtain analytical form of the following integral in Eq.~\eqref{defn1}
\begin{gather}
    \sum \limits_{\lambda} \int \limits_{B.Z.} \approx \sum \limits_{\lambda} \int \limits^{\infty}_0 dk \int \limits_{0}^{2\pi} d \theta
    = \sum \limits_{\lambda} \int \int  \ ,
\end{gather}
where the upper limit of the $k$ integral is extended to $\infty$ and we used $\theta_k=\theta+\alpha_{ll^\prime}$ with $\alpha_{ll^\prime}$ the angle which ${\bf r}_{ll'}$ makes with the positive $k_x$-axis. This leads to
\begin{eqnarray*}
G_{\texttt{AA}}&=&\frac{{2\cal A}}{(2\pi)^2} \cos\left({\bf K} {\bf r}_{ll' } \right) \int \int    \frac{E^2-\epsilon_k^2\sin^2\phi}{E\left(E^2-\epsilon_k^2\right)} \texttt{e}^{i{\bf k}{\bf r}_{ll'}} \ ,
\nonumber\\
G_{\texttt{BB}}&=& \frac{{2\cal A}}{(2\pi)^2} \cos\left({\bf K} {\bf r}_{ll' } \right) \int \int     \frac{E^2}{E^2-\epsilon_k^2} \texttt{e}^{i{\bf k}{\bf r}_{ll'}} \ ,
\nonumber\\
G_{\texttt{CC}}&=&\frac{{2\cal A}}{(2\pi)^2}\cos\left({\bf K} {\bf r}_{ll' } \right) \int \int     \frac{E^2-\epsilon_k^2\cos^2\phi}{E\left(E^2-\epsilon_k^2\right)} \texttt{e}^{i{\bf k}{\bf r}_{ll'}} \ ,
\nonumber\\
G_{\texttt{AB}}&=& \frac{{\cal A}}{(2\pi)^2} \left[ \texttt{e}^{i({\bf K}{\bf r}_{ll'}-\alpha_{ll^\prime})} \int \int  \frac{ \epsilon_k\cos\phi }{ E^2-\epsilon_k^2 } \texttt{e}^{i\left( {\bf k}{\bf r}_{ll'}-\theta \right)} \right. 
\nonumber\\
&-&\left. \texttt{e}^{-i({\bf K}{\bf r}_{ll'}- \alpha_{ll^\prime}) )} \int \int \frac{ \epsilon_k\cos\phi  }{ E^2-\epsilon_k^2 } \texttt{e}^{i\left( {\bf k}{\bf r}_{ll'}+\theta \right)} \right] \ ,
\nonumber\\
G_{\texttt{AC}}&=& \frac{{\cal A}}{(2\pi)^2}
\left[ \texttt{e}^{i({\bf K}{\bf r}_{ll'}-2 \alpha_{ll^\prime})} \int \int  \frac{ \epsilon_k^2\sin(2\phi )}{2E( E^2-\epsilon_k^2) } \texttt{e}^{i\left( {\bf k}{\bf r}_{ll'}-2\theta \right)} \right. 
\nonumber\\
&+&\left.  \texttt{e}^{-i({\bf K}{\bf r}_{ll'}-2 \alpha_{ll^\prime})} \int \int \frac{ \epsilon_k^2\sin(2\phi )}{2E( E^2-\epsilon_k^2) } \texttt{e}^{i\left( {\bf k}{\bf r}_{ll'}+2\theta \right)} \right] ,
\nonumber\\
G_{\texttt{BC}}&=&  \frac{{\cal A}}{(2\pi)^2} \left[ \texttt{e}^{i({\bf K}{\bf r}_{ll'}-\alpha_{ll^\prime})} \int \int \frac{ \epsilon_k\sin\phi }{ E^2-\epsilon_k^2 }\texttt{e}^{i\left( {\bf k}{\bf r}_{ll'}-\theta \right)} \right.
\nonumber\\
&-&\left.  \texttt{e}^{-i({\bf K}{\bf r}_{ll'}-\alpha_{ll^\prime})} \int \int \frac{ \epsilon_k\sin\phi  }{ E^2-\epsilon_k^2 }\texttt{e}^{i\left( {\bf k}{\bf r}_{ll'}+\theta \right)} \right] \ ,
\label{eqn:GGG3}
\end{eqnarray*}
The above expressions can also be written in the form
\begin{eqnarray}
G_{\texttt{AA}}&=& \cos\left({\bf K}{\bf r}_{ll'}\right) F_{\texttt{AA}}({\bf r}_{ll'},E;\phi)
\nonumber\\
G_{\texttt{BB}}&=&  G_{\texttt{AA}}({\bf r}_{ll'},E;\phi=0)
\nonumber\\
G_{\texttt{CC}}&=& G_{\texttt{AA}}({\bf r}_{ll'},E;\phi+\pi/2)
\nonumber\\
G_{\texttt{AB}}&=& \sin\left({\bf K}{\bf r}_{ll'}-\alpha_{ll^\prime}  \right) F_{\texttt{AB}}({\bf r}_{ll'},E;\phi)
\nonumber\\
G_{\texttt{AC}}&=& \cos\left({\bf K}{\bf r}_{ll'}-2\alpha_{ll^\prime}  \right) F_{\texttt{AC}}({\bf r}_{ll'},E;\phi)\nonumber\\
G_{\texttt{BC}}&=& \sin\left({\bf K}{\bf r}_{ll'}-\alpha_{ll^\prime}  \right) F_{\texttt{BC}}({\bf r}_{ll'},E;\phi) \  ,
\label{theGs}
\end{eqnarray}
if we define the following auxiliary quantities given by the Hankel transforms
\begin{eqnarray}
F_{\texttt{AA}}&=&  \  \frac{{\cal A}}{\pi}  \int_0^\infty dk \ k \ J_0\left(k r_{ll'}\right)     \left[\frac{E^2-\epsilon_k^2\sin^2\phi}{E\left(E^2-\epsilon_k^2\right)}\right]
\nonumber\\   
&=&    -\   \epsilon^2  \left( \frac{{\cal A}}{\pi a^2 E}\right)    K_0(-i r\epsilon)\cos^2\phi \ ,
\nonumber\\
F_{\texttt{AB}}&=&-\
\frac{{\cal A}}{\pi}  \int_0^\infty   dk \ k \ J_1\left(k r_{ll'}\right)     \left(\frac{\epsilon_k\cos\phi}{ E^2-\epsilon_k^2}\right) 
\nonumber\\
&=&    -i  \   \epsilon^2    \left( \frac{{\cal A}}{\pi a^2 E}\right)   \cos(\phi) \  K_1(-i r\epsilon) \ ,
\nonumber\\
F_{\texttt{AC}}&=&
\frac{{\cal A}}{\pi}  \int_0^\infty  dk \ k \ J_2\left(k r_{ll'}\right)     \left[\frac{\epsilon_k^2\sin(2\phi)}{2E( E^2-\epsilon_k^2)}\right]
\nonumber\\
&=&     \frac{1}{2}   \epsilon^2    \left( \frac{{\cal A}}{\pi a^2 E}\right)   \sin(2\phi) \  K_2(-i r\epsilon) \ ,
\nonumber\\
F_{\texttt{BB}}&=&  F_{\texttt{AA}}(r_{ll'},E;\phi=0) \ ,
\nonumber\\
F_{\texttt{CC}}&=& F_{\texttt{AA}}(r_{ll'},E;\phi+\pi/2) \ ,
\nonumber\\
F_{ \texttt{BC} }&=& -\
\frac{{\cal A}}{\pi}  \   \int_0^\infty dk \ k \ J_1\left(k r_{ll'}\right)     \left[\frac{\epsilon_k\sin(\phi)}{ E^2-\epsilon_k^2}\right]
\nonumber\\
&=&     -i\     \epsilon^2   \left( \frac{{\cal A}}{\pi a^2 E}\right)   \sin(\phi) \  K_1(-i r\epsilon)
 \  .
\label{theFs}
\end{eqnarray}
Here, we employed a well known identity
\begin{equation}
\int_0^\infty dx\ \frac{x^{n+1}}{x^2+C^2}J_n(xR)=C^n\ K_n(-CR) \ ,\nonumber
\end{equation}
 where $K_n(x)$  ($n=0,1,2,\cdots$) is a modified Bessel function of the second kind. For convenience, we used the notation 
$\epsilon=E/\gamma$, where $\gamma=\hbar v_F/a$, and $r=|{\bf r}_{l}-{\bf r}_{l^\prime}|  /a$ where $a$ is the separation between the $ \texttt{A}$ and $\texttt{B}$ sites of the hexagonal lattice. Also, $v_F$ is the Fermi velocity.
Together Eqs.\eqref{theGs} and \eqref{theFs} yield the desired final expression.

\section{\label{AP:3} Derivation of Eq.~(\ref{EQ:main})}

The integration over $k_y$ in Eq.\eqref{EQ:suceptibility} can be performed analytically using 
\begin{gather}
 \sum \limits_{k_y} = \frac{\mathcal{A}}{2 \pi} \int \limits_{-\infty}^{\infty} \frac{d\left[Y+\left(x_{l'}+x_{l}\right)/2+i\left(y_{l'}-y_{l}\right)/2\right]}{2\pi l_H^2} \ ,
 \end{gather}
then the expression for the wave-functions overlap becomes
\begin{gather}
 \Phi_{n+\lambda \mu}^{n+\lambda \nu} \left({\mathbf{r}_l,\mathbf{r}_{l'}}\right)=\\
 \notag
 =\sum \limits_{k_y} \phi_{n+\lambda \mu, k_y}\left({x_l}\right) \phi_{n+\lambda \nu, k_y}\left({x_l}\right)
  \texttt{e}^{-i k_y \left({y_{l}-y_{l'}}\right)}\\
	\notag
 =
\frac{\mathcal{A}}{2 \pi}
\frac{\exp\left[{-\frac{r^2_{ll'}}{4} - i \frac{\left({x_{l}+x_{l'}}\right)\left({y_{l}-y_{l'}}\right)}{2 l_H^2} }\right]}{2 \pi^{3/2} l_H^2 \sqrt{2^{n + \lambda \mu} \left({n+\lambda \mu}\right)!} \sqrt{2^{n + \lambda \nu} \left({n+\lambda \nu}\right)!}}  \\
\notag\times
\int \limits_{-\infty}^{\infty} dy \texttt{e}^{-y^2} H_{n+\lambda \mu} \left({x-y}\right)  H_{n+\lambda \nu} \left({z-y}\right) \  ,
\end{gather}
where $Y=k_y l_H^2$; $y=Y/l_H$; $x = \frac{\left({x_{l}-x_{l'}}\right)+ i \left({y_{l}-y_{l'}}\right)}{2 l_H} = \frac{r_{ll'}}{2} \exp  \left({i \alpha_{ll'}}\right)$
; $z = \frac{\left({x_{l'}-x_{l}}\right)+ i \left({y_{l}-y_{l'}}\right)}{2 l_H} = -\frac{r_{ll'}}{2} \exp\left({ - i \alpha_{ll'}}\right)$ and
$r_{ll'}=2|x|=2|z|$.

\medskip
\par
Now, let us use the following integral relation
\begin{gather}
   \int \limits_{-\infty}^{\infty} dy\   \texttt{e}^{-y^2} H_{n+\lambda \mu} \left({x-y}\right)  H_{n+\lambda \nu} \left({z-y}\right)\\
   \notag
   = \sqrt{\pi} 2^n
   \begin{cases}
   2^{\lambda \nu} \left({n+\lambda \mu}\right)! z^{\lambda \left({\nu - \mu}\right)} L^{\lambda \left({\nu - \mu}\right)}_{n + \lambda \mu} \left({\frac{r^2_{ll'}}{2}}\right) &   ;\  \lambda \mu \leq \lambda \nu\\
   2^{\lambda \mu} \left({n+\lambda \nu}\right)! x^{\lambda \left({\mu - \nu}\right)} L^{\lambda \left({\mu - \nu}\right)}_{n + \lambda \nu} \left({\frac{r^2_{ll'}}{2}}\right) &    ;\   \lambda \mu > \lambda \nu \  .
   \end{cases}
\end{gather}
Including the flat band to the overlap function, we finally obtain  
\begin{gather}\label{Eq:Phi}
  \Phi_{n+\lambda \mu}^{n+\lambda \nu} \left({s; \mathbf{r}_l,\mathbf{r}_{l'}}\right)=\frac{\mathcal{A}}{\left({2 \pi l_H}\right)^2} \tilde{\Phi}_{n+\lambda \mu}^{n+\lambda \nu} \left({s;\mathbf{r}_l,\mathbf{r}_{l'}}\right), \\
  \notag
  \tilde{\Phi}_{n+\lambda \mu}^{n+\lambda \nu} \left({s,\mathbf{r}_l,\mathbf{r}_{l'}}\right)=
\exp\left[{-\frac{r^2_{ll'}}{4} - i \frac{\left({x_{l}+x_{l'}}\right)\left({y_{l}-y_{l'}}\right)}{2 l_H^2} }\right] \times\\
\notag
   \begin{cases}
  \sqrt{\frac{2^{\lambda \nu} (n+\lambda\mu)!}{2^{\lambda\mu}(n+\lambda\nu)!}}z^{\lambda(\nu-\mu)}
  L^{\lambda \left({\nu - \mu}\right)}_{n + \lambda \mu} \left({\frac{r^2_{ll'}}{2}}\right) &  ;\mbox{for}\   \lambda \mu \leq \lambda \nu\\
   \sqrt{\frac{2^{\lambda \mu} (n+\lambda\nu)!}{2^{\lambda\nu}(n+\lambda\mu)!}}x^{\lambda(\mu-\nu)}
    L^{\lambda \left({\mu - \nu}\right)}_{n + \lambda \nu} \left({\frac{r^2_{ll'}}{2}}\right) &   ;\   \lambda \mu > \lambda \nu\\
   0 &    ;\   n+\texttt{min}\left({\lambda \mu,\lambda \nu}\right) <0\\
    L^{0}_{0} \left({\frac{r^2_{ll'}}{2}}\right) &    ;\   n=0,s=0 \ .
   \end{cases}
\end{gather}
Substituting  Eqs.~\eqref{Eq:psi} and \eqref{Eq:Phi} into Eq.~\eqref{Eq:Psi} and the resulting equation into Eq.~\eqref{EQ:suceptibility}, we finally obtain Eq.~\eqref{EQ:main}.

 \end{document}